%% file: chic0BR.tex
\documentclass[11pt]{article}
\usepackage{graphicx}

\newcommand{\BABARPubYear}    {02}

\newcommand{\BABARConfNumber} {22}
\newcommand{\SLACPubNumber} {9316}

\input pubboard/babarsym
\input symbols

\long\def\inst#1{\par\nobreak\kern 4pt\nobreak
    {\it #1}\par\vskip 10pt plus 3pt minus 3pt}

\begin{document}
{\pagestyle{empty}

\begin{flushright}
\babar-CONF-\BABARPubYear/\BABARConfNumber \\
SLAC-PUB-\SLACPubNumber \\
July 2002 \\
\end{flushright}

\par\vskip 5cm

\begin{center}
\Large \bf Measurement of the Branching Fraction for \boldmath \Bpm\to\chiczero\Kpm
\end{center}
\bigskip

\begin{center}
\large The \babar\ Collaboration\\
\mbox{ }\\
\today
\end{center}
\bigskip \bigskip

\begin{center}
\large \bf Abstract
\end{center}

\input abstract

\vfill
\begin{center}
Contributed to the 31$^{st}$ International Conference on High Energy Physics,\\ 
7/24---7/31/2002, Amsterdam, The Netherlands
\end{center}

\vspace{1.0cm}
\begin{center}
{\em Stanford Linear Accelerator Center, Stanford University, 
Stanford, CA 94309} \\ \vspace{0.1cm}\hrule\vspace{0.1cm}
Work supported in part by Department of Energy contract DE-AC03-76SF00515.
\end{center}

\newpage
}

\input pubboard/authors_ICHEP2002.tex

\section{Introduction}
\label{sec:Introduction}
The \chiczero\ state corresponds, in spectroscopic notation, to the $1^3P_0$ state of the charmonium spectrum ($J^{PC}=0^{++}$), whose mass and width are measured to be $(3415.1\pm0.8)\mevcc$ and $(14.9^{+2.6}_{-2.3})\mev$, respectively \cite{ref:pdg2002}.

In the framework of the factorization approximation, the production rate for \BToChiczeroK\ decays vanishes when the assumption is made that this process occurs through the decay of a \b quark into a color-singlet \ccbar pair at short distances \cite{ref:kuhn}. It has been suggested \cite{ref:bodwin, ref:beneke} that the ``color-octet mechanism'' could enhance such decays via the emission of a soft gluon. The branching fraction for \B \to \chiczero \kaon\ would then be expected to be comparable in magnitude to that of the ``color-singlet-allowed'' decay mode \B \to \chicone \kaon.

Belle recently reported the first evidence for the \BchToChiczeroKch\ decay \cite{ref:belleChic0}. On a sample of $31.3 \times 10^6$ \BB\ events, Belle measured $\BR(\BchToChiczeroKch) = (6.0^{+2.1}_{-1.8} \pm 1.1)\times 10^{-4}$; the \chiczero\ meson was reconstructed through its decay to the \pipi\ mode. Previously CLEO had reported an upper limit of $\BR(\BchToChiczeroKch) < 4.8\times 10^{-4}$ at 90$\%$ C.L. \cite{ref:cleoChic0}.

The results of a similar analysis performed on \babar\ data are reported in this paper; both \chiczeroToPipi\ and \chiczeroToKK\ decay modes have been used for the measurement of \BR(\BchToChiczeroKch); the branching fractions of \chiczeroToPipi\ and \chiczeroToKK\ are $(5.0\pm0.7)\times10^{-3}$ and $(5.9\pm0.9)\times10^{-3}$ respectively \cite{ref:pdg2002}. The choice of these two modes was motivated by their cleaner signature compared to other higher multiplicity hadronic modes showing larger branching fractions.

\section{The \babar\ detector and dataset}
\label{sec:babar}
The data used in this analysis were collected with the \babar\ detector at the \pep2\ $e^+e^-$ storage ring. The data sample we used corresponds to about 70 \invfb of integrated luminosity collected at  the \FourS\ resonance (``on-resonance'') between October 1999 and May 2002. The collider is operated in an asymmetric mode, that is with beams of unequal energies, so that the center of mass is ``boosted'' (with $\beta\gamma=0.55$) along the direction of the electron beam.

The \babar\ detector is fully described elsewhere \cite{ref:babar}. It consists of a charged particles tracking system, a Cherenkov detector (DIRC) for particle identification, an electromagnetic calorimeter and a system for muon and \KL\ identification. The tracking system consists of a 5-layer, double-sided silicon vertex tracker and a 40-layer drift chamber (filled with a mixture of helium and isobutane), both in a 1.5 T magnetic field supplied by a superconducting solenoidal magnet. The DIRC is a unique imaging Cherenkov detector relying on total internal reflection in the radiator. The electromagnetic calorimeter consists of 6580 CsI(Tl) crystals. The iron flux return is segmented and instrumented with resistive plate chambers for muon and \KL\ identification.

\section {Analysis method}

A preliminary selection for hadronic events required the presence of at least three charged tracks, the ratio between the 2nd and 0th order Fox-Wolfram moments to be lower than 0.5 and the total energy of all the charged and neutral particles in the event to be greater than 4.5 \gev. Furthermore, at least one track identified as kaon was required to have a momentum greater than $900 \mevc$ in the center of mass frame.

\Bpm\ meson candidates were reconstructed by combining a track identified as a charged kaon with a \chiczero\ candidate. The kaon four-momentum was required to be loosely consistent with the kinematics of the decay; the \chiczero\ meson was reconstructed from a pair of charged tracks identified as both pions or both kaons. The typical efficiency for the kaon selectors used is in the range $70\%$ to $90\%$, depending on momentum, while the typical probability for a pion to be misidentified as a kaon is below $5\%$. All these tracks were explicitly required to have polar angles lying in the region $0.35 \rad < \theta < 2.54 \rad$, to have at least 12 hits in the drift chamber and a transverse momentum with respect to the beam direction greater than $100 \mevc$. In addition, tracks identified as decay products of $\KS\to\pipi$, $\eta \to \pi^+\pi^-\pi^0$, $\Lambda \to p\pi$ or tracks from \g\ conversions were rejected. The \chiczero\ candidates were then defined as those for which the \pipi(\Kp\Km) invariant mass lies in the region $3.32 \gevcc < \mChiczero\ < 3.50 \gevcc$.

In order to reject the large background coming from continuum events, a Fisher discriminant has been used, built from a linear combination of eleven event shape or \B\ kinematics related quantities. The coefficients were determined, separately for the two modes, by maximizing the achieved separation between signal and continuum background (both from samples of events simulated with Monte Carlo techniques). 

The selection of \B\ candidates relied on the kinematic constraints given by the \FourS\ initial state. Two largely uncorrelated variables were used: the beam-energy substituted mass, $\mes = \sqrt{E^{*2}_{beam} - |\mathbf{p}^*|^2}$, and $\DEev = E^*-\sqrt{m^2_B+|\mathbf{p}^*|^2}$, where $E^*_{beam}=\sqrt{s}/2$ is the expected \B\ energy in the \FourS\ rest frame, $E^*$ and $\mathbf{p}^*$ are respectively the energy and vector momentum of the \B\ candidate in the \FourS\ rest frame, and $m_B$ is the mass of the \Bpm\ meson. In cases where multiple \B\ candidates were present in the same event, only the one with the smallest absolute value of \DEev\ was retained.

The values of the cuts for the Fisher discriminant, \mes\ and \DEev\ were determined by an optimization procedure aimed at maximizing the value of $S/\sqrt{S+B}$. The number $S$ of signal candidates and $B$ of background events surviving the cuts were estimated on samples of simulated events and on data from the \DEev\ ``sidebands''  of the \mes\ vs. \DEev\ plane, respectively. The sidebands were defined by $5.2 \gevcc < \mes < 5.3 \gevcc$, $0.1 \gev < |\DEev| < 0.2 \gev$. The signal and background samples were normalized to each other assuming the value measured by Belle for \BR(\BchToChiczeroKch) and the world average for \BR(\chiczeroToPipi) and \BR(\chiczeroToKK) \cite{ref:pdg2002}. 

The signal region in the \mes\ vs. \DEev\ plane was defined by  $-45 \mev < \DEev < 35 \mev$, $\mes>5.275 \gevcc$ for the \chiczeroToPipi\ mode and by $-70 \mev < \DEev < 60 \mev$, $\mes>5.2735 \gevcc$ for the \chiczeroToKK\ mode. To account for a shift in \DEev\ between Monte Carlo and data, observed on samples of $\B\to\Dz\pi$ events with a \Kpm\pipi\ final state, a correction of +7 \mev\ has been applied to \DEev\ on data for both modes. Table~\ref{tab:sigBox} summarizes these cuts and the number of selected events in the signal box.

\begin{table}[!htb]
\caption{Signal box definition (including the region in \mChiczero) and number of selected events for the two modes.}
\begin{center}
\begin{tabular}{|c|c c c|c|} \hline
Mode & \mChiczero\ (\gevcc) & \DEev\ (\mev) & \mes\ (\gevcc) & $N_{\rm sel}$ \\
\hline\hline
\chiczeroToPipi & $3.32 \div 3.50$ & $-45 \div 35$ & $>5.2750$ & 83 \\
\chiczeroToKK & $3.32 \div 3.50$ & $-70 \div 60$ & $>5.2735$ & 84 \\
\hline
\end{tabular}
\end{center}
\label{tab:sigBox}
\end{table}

In order to remove the contamination from \B\ decays to $D$ and \Dstar\ modes, an explicit veto on several of these modes has been applied: \B\ candidates were rejected if at least one of their decay products also contributed to the reconstruction of a \B\ in one of the veto modes, with $|\DEev| < 30 \mev$ and $\mes>5.27 \gevcc$. 

Some residual contamination was still observed from modes with a \Dz\ meson decaying to $K^-\pi^+$ or a $\phi$ meson decaying to \KpKm, for \Bpm\to\chiczero(\to\pipi)\Kpm\ and \Bpm\to\chiczero(\to\KpKm)\Kpm, respectively. This background was suppressed by rejecting all events for which the invariant mass of the pair formed by the fast kaon with the oppositely charged pion (kaon) was compatible with the \Dz\ ($\phi$) mass.

The main source of non-combinatorial background remaining after the selection described comes from non resonant \B\ decays with the same final state as the signal: \Bpm\to\Kpm\pipi\ and \Bpm\to\Kpm\KpKm. The fraction of such events surviving the set of cuts detailed above is $1.4\times 10^{-2}$ and $3.5\times 10^{-2}$, respectively; these estimates however could not be used for a reliable evaluation of the expected contamination, since the associated branching fractions are not well known. These modes are expected to peak in \mes\ and \DEev\ (what will be referred to as ``peaking background'' in the following), while the distribution of \mChiczero\ is expected to be flat: this was used to separate their contribution from the signal by means of a fit to the data, as will be described in the following section.

The background from mis-reconstructed \chiczero\ decays to other modes has been estimated to be negligible with respect to the other background sources for both the \pipi\ and the \KpKm\ modes.

The overall selection efficiency, estimated from Monte Carlo, is $(31.2\pm0.6\pm1.6)\%$ for the \chiczeroToPipi\ mode and $(24.1\pm0.5\pm1.1)\%$ for the \chiczeroToKK\ mode.  In both cases, the first error is statistical and the second reflects the systematic uncertainties. These are mainly due to differences between data and Monte Carlo in the track reconstruction and particle identification efficiencies; a smaller contribution comes from differences in the efficiency of the selection cuts, in particular the cut on the Fisher discriminant.

\section{Extraction of the signal yield}
\label{sec:Yield}

The extraction of the signal yield was complicated by the large fraction of background events (both combinatorial and peaking background) in the selected sample. \B\ decays were separated from the combinatorial background by means of the two kinematic variables \mes\ and \DEev; the \mChiczero\
 distribution was used to separate the signal and the peaking background contributions.

The number of signal events was extracted by fitting the \mChiczero\ distribution for events in the signal box with a generic background component (combinatorial + peaking) and a signal component. An unbinned maximum likelihood fit was used.

Background events were assumed to have a uniform \mChiczero\ distribution in the narrow region selected for this variable. This was confirmed by looking at the \mChiczero\ distribution for data events away from the (\mes, \DEev) signal box, and for \Bpm\to\Kpm\pipi, \Bpm\to\Kpm\KpKm\ Monte Carlo events. The signal component was modelled by a non relativistic Breit-Wigner function convoluted with a Gaussian function. In the fit the mass and the total width of the \chiczero\ were fixed to the expected values; the width of the Gaussian function was determined from Monte Carlo. The reliability of the fit was checked on a ``cocktail'' sample containing known amounts of combinatorial background, peaking background and signal events.

The resulting signal yields are $N_{\rm sig}(\chiczeroToPipi) = 25.4 \pm 8.2$ and $N_{\rm sig}(\chiczeroToKK)  = 27.0 \pm 8.0$. The breakdown of the background contributions, as estimated by combining the above results with fits to the \mes\ distributions, is $N_{\rm bkg}(\chiczeroToPipi) = 26.1$(combinatorial)$ + 31.5$(peaking) and $N_{\rm bkg}(\chiczeroToKK) = 20.8$(combinatorial)$ + 36.2$(peaking). These values refer to the whole region $3.32 \gevcc < \mChiczero\ < 3.5 \gevcc$. 

\begin{figure}[!hb]
\begin{center}
\includegraphics[height=15cm]{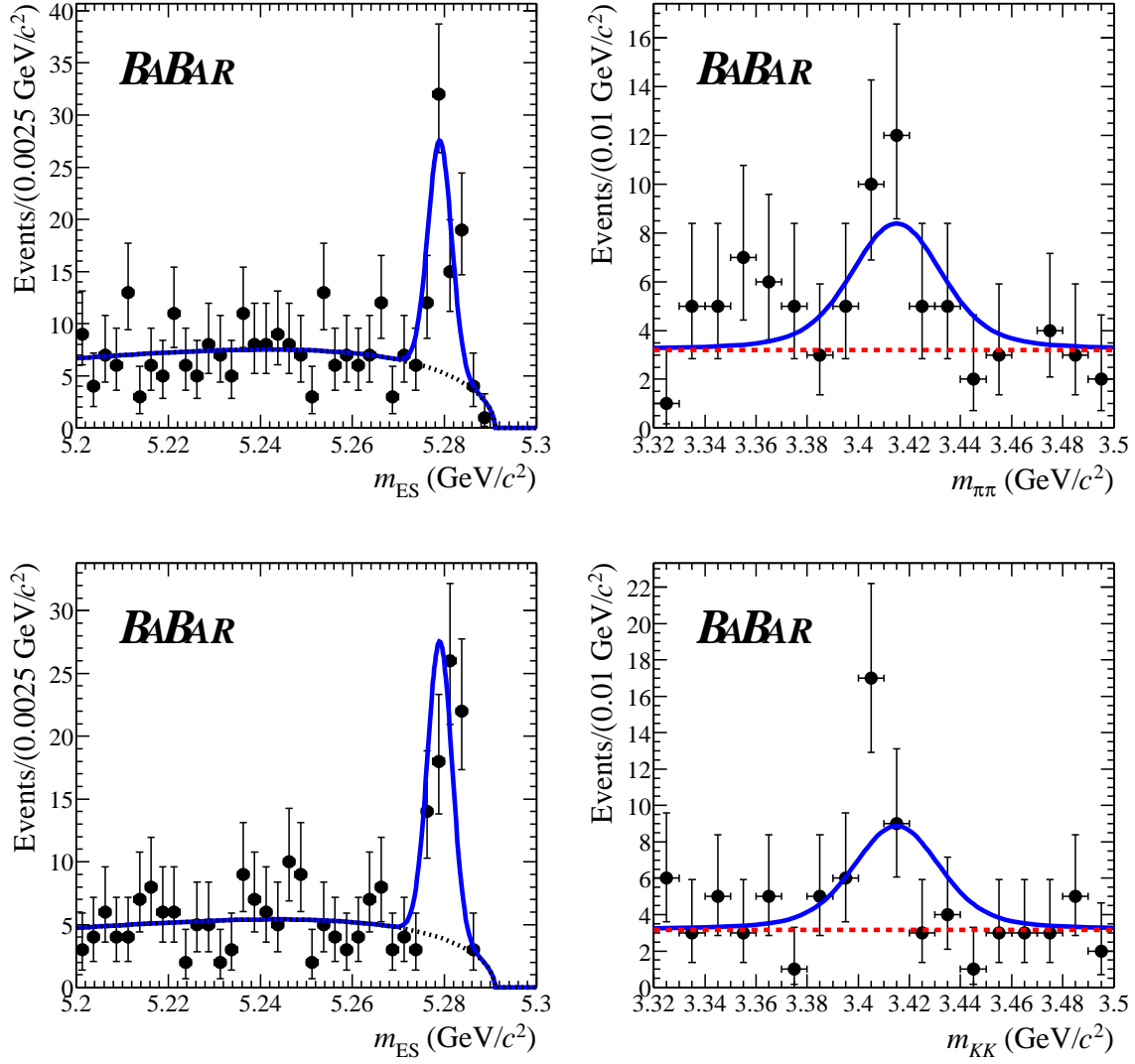}
\caption{Extraction of the signal yield for \chiczeroToPipi\ (top) and \chiczeroToKK\ (bottom).
Left: distribution of the \mes\ variable for events selected in the \DEev\ signal region. Superimposed is the result of the fit to a Gaussian \B\ signal component (solid line) + an Argus function for the continuum background (dotted line).
Right: distribution of the recoil invariant mass \mChiczero\ for events in the signal box. Superimposed is the result of the fit to a Breit-Wigner function convoluted with a Gaussian function (solid line) + a flat background component (dotted line).}
\label{fig:fit_pipi_KK}
\end{center}
\end{figure}

Figure \ref{fig:fit_pipi_KK} shows the \mes\ distribution for events in the \DEev\ signal region and the \mChiczero\ distribution for events in the signal box, for the two modes considered. Superimposed are the fit results. Comparison of the \mes\ and \mChiczero\ plots reveals the presence of a substantial component of non-\chiczero\ events with a peaking structure in \mes\ in both cases. Furthermore, the enhancements seen in the \mChiczero\ distribution appear shifted to lower values with respect to the expected \chiczero\ mass. The distortion of the \mChiczero\ distributions could originate from the presence of interference effects with non-resonant 3-body decays; this effect has not been taken into account in this analysis. As a check, we repeated the fit by allowing the \chiczero\ mass to float, and obtaining for it values shifted by about -6 \mevcc\ (for \chiczeroToPipi) and -10 \mevcc\ (for \chiczeroToKK) with respect to the published PDG values. Results for the yield were compatible well within 1 $\sigma$ with those extracted with the mass fixed.

As a way to estimate systematic effects related to the fitting procedure, an alternate method was employed. There, the \mes\ and \mChiczero\ distributions were simultaneously fit for all events in the \DEev\ signal region. In this case the likelihood function contained three terms: one for combinatorial background, one for the peaking background and one for the signal. The \mes\ distribution for the peaking background component was modelled with a Gaussian with the same resolution as the signal, while the combinatorial component was fit to the usual empirical Argus shape. Both backgrounds were assigned uniform distributions in \mChiczero. This fit yields  $N_{\rm sig}(\chiczeroToPipi) = 23.4 \pm 8.5$ and $N_{\rm sig}(\chiczeroToKK)  = 29.7 \pm 8.5$. The absolute difference in the signal yield with respect to the previous method is taken as a global systematic error.

\section{Branching fraction measurement}
\label{sec:Physics}

The branching fractions are obtained as $\BR = \frac{N_{\rm sig}}{\epsilon \times N_{\Bpm}}$, where $\epsilon$ denotes the overall signal efficiency and $N_{\Bpm}$ is the total number of \Bpm\ mesons produced in the data sample considered.

$N_{\Bpm}$ is obtained from the measured number of \BB\ pairs, $N_{\BB} = (75.7\pm1.1) \times 10^6$, assuming $\BR(\FourS \to \BpBm) = 50\%$.

The results for the product of the branching fractions for the two modes are reported in Table~\ref{tab:BR}. The systematic errors reported there combine the uncertainty on the efficiency evaluation with the error on $N_{\BB}$ and that associated with the fitting procedure.

In order to extract the value of the branching fraction for \BchToChiczeroKch, we used the values reported by the PDG \cite{ref:pdg2002} for \BR(\chiczeroToPipi) and \BR(\chiczeroToKK), to obtain:

\begin{center}
$\BR (\BchToChiczeroKch) = (2.15 \pm 0.69 \stat \pm 0.21\syst \pm 0.30 \syst) \times 10^{-4}$ 
\end{center}

for \chiczeroToPipi\ and 

\begin{center}
$\BR (\BchToChiczeroKch) = (2.51 \pm 0.74 \stat \pm 0.28 \syst \pm 0.38 \syst) \times 10^{-4}$
\end{center}

for \chiczeroToKK; here the second systematic error accounts for the uncertainty on the \chiczero\ decay branching fractions. The two measurements were combined to obtain:

\begin{center}
$\BR (\BchToChiczeroKch) = (2.4 \pm 0.7) \times 10^{-4}$
\end{center}

When combining the errors, we treated the systematic uncertainties arising from efficiency corrections, from the fitting procedure and from the number of \BB\ pairs determination as fully correlated, while statistical errors have been treated as uncorrelated, since no events belong simultaneously to both the samples. To combine the errors coming from the measurement of the \chiczero\ branching fractions, we only used the values published by the BES Collaboration \cite{ref:BES_chic0} and treated the errors as fully correlated.

The measurement obtained for \BR(\BchToChiczeroKch) is of the same order of magnitude as that for \BchToChiconeKch\ ($(6.5 \pm 1.1) \times 10^{-4}$ from \cite{ref:pdg2002}), thus supporting the treatment of charmonium production in \B\ decays with a QCD factorization approach including non-perturbative color-octet contributions. 

\begin{table}[!htb]
\caption{Branching fractions from the two \chiczero\ decay modes and their combined value. The meaning of the quoted errors is explained in the text.}
\begin{center}
\begin{tabular}{|c|c|c|} \hline
Mode & $\BR(\BchToChiczeroKch)\times\BR(\chiczero\to f)\ (10^{-6})$ & $\BR(\BchToChiczeroKch)\ (10^{-4})$ \\
\hline\hline
\chiczeroToPipi & $1.08\pm0.35\stat \pm0.10\syst$ & $2.15 \pm 0.69 \stat \pm 0.21\syst \pm 0.30 \syst$ \\
\chiczeroToKK & $1.48\pm0.44\stat \pm0.17\syst$ & $2.51 \pm 0.74 \stat \pm 0.28 \syst \pm 0.38 \syst$ \\
\hline
Combined & & $2.4 \pm 0.7$\\
\hline
\end{tabular}
\end{center}
\label{tab:BR}
\end{table}

\section{Acknowledgments}
\label{sec:Acknowledgments}

\input pubboard/acknowledgements

\end{document}

%% file: symbols.tex
\def\BToChiczeroK {\ensuremath{\B \to \chiczero \kaon}}
\def\BchToChiczeroKch {\ensuremath{\Bpm \to \chiczero \Kpm}} 
\def\BchToChiconeKch {\ensuremath{\Bpm \to \chicone \Kpm}} 
\def\chiczeroToPipi {\ensuremath{\chiczero \to \pipi}}
\def\chiczeroToKK {\ensuremath{\chiczero \to \KpKm}}

\def\DEev {\ensuremath{\DeltaE_{\rm ev}}}
\def\mChiczero {\ensuremath{m_{\chiczero}}}

%% file: abstract.tex
We present preliminary results for the measurement of the branching fraction of the decay $B^\pm\to\chi_{c0}K^\pm$ from a sample of 75 million $B\kern 0.18em\overline{\kern -0.18em B}{}$ pairs collected by the \mbox{\slshape B\kern-0.1em{\smaller A}\kern-0.1em B\kern-0.1em{\smaller A\kern-0.2em R}} detector at the \mbox{PEP-II} asymmetric-energy $B$ Factory at SLAC. The $\chi_{c0}$ meson is reconstructed through its two-body decays to $\pi^+\pi^-$  and $K^+ \kern -0.16em K^-$. We measure $\cal B$$(B^\pm\to\chi_{c0}K^\pm)\times$$\cal B$$(\chi_{c0}\to\pi^+\pi^-) = (1.08\pm0.35\mathrm{(stat)} \pm0.10\mathrm{(syst)})\times10^{-6}$ and $\cal B$$(B^\pm\to\chi_{c0}K^\pm)\times$$\cal B$$(\chi_{c0}\to K^+ \kern -0.16em K^-) = (1.48\pm0.44\mathrm{(stat)} \pm0.17\mathrm{(syst)})\times 10^{-6}$. Using the known values for the $\chi_{c0}$ decays branching fractions, we combined these results to obtain $\cal B$$(B^\pm\to\chi_{c0}K^\pm) = (2.4\pm0.7)\times 10^{-4}$.

%% file: pubboard/authors_ICHEP2002.tex
\begin{center}
\small

The \babar\ Collaboration,
\bigskip

B.~Aubert,
D.~Boutigny,
J.-M.~Gaillard,
A.~Hicheur,
Y.~Karyotakis,
J.~P.~Lees,
P.~Robbe,
V.~Tisserand,
A.~Zghiche
\inst{Laboratoire de Physique des Particules, F-74941 Annecy-le-Vieux, France }
A.~Palano,
A.~Pompili
\inst{Universit\`a di Bari, Dipartimento di Fisica and INFN, I-70126 Bari, Italy }
% ...

\end{center}\newpage

%% file: pubboard/acknowledgements.tex
We are grateful for the 
extraordinary contributions of our \pep2\ colleagues in
achieving the excellent luminosity and machine conditions
that have made this work possible.
The success of this project also relies critically on the 
expertise and dedication of the computing organizations that 
support \babar.
The collaborating institutions wish to thank 
SLAC for its support and the kind hospitality extended to them. 
This work is supported by the
US Department of Energy
and National Science Foundation, the
Natural Sciences and Engineering Research Council (Canada),
Institute of High Energy Physics (China), the
Commissariat \`a l'Energie Atomique and
Institut National de Physique Nucl\'eaire et de Physique des Particules
(France), the
Bundesministerium f\"ur Bildung und Forschung and
Deutsche Forschungsgemeinschaft
(Germany), the
Istituto Nazionale di Fisica Nucleare (Italy),
the Research Council of Norway, the
Ministry of Science and Technology of the Russian Federation, and the
Particle Physics and Astronomy Research Council (United Kingdom). 
Individuals have received support from 
the A. P. Sloan Foundation, 
the Research Corporation,
and the Alexander von Humboldt Foundation.